\documentclass[aps,twocolumn,showpacs,twoside]{revtex4}
\usepackage{amsmath,amssymb,epsfig}
\begin{document}
\title{Reactive Hall constant of Strongly Correlated Electrons}
\author{P. Prelov\v sek}
\email{peter.prelovsek@ijs.si}
\affiliation{Faculty of Mathematics and Physics, University of
Ljubljana, 1000 Ljubljana, Slovenia, and J. Stefan Institute, 1000
Ljubljana, Slovenia }
\author{X. Zotos}
\email{Xenophon.Zotos@epfl.ch}
\affiliation{Institut Romand de Recherche Num\'erique en Physique des
Mat\'eriaux (EPFL-PPH), CH-1015 Lausanne, Switzerland }
\date{\today}
\begin{abstract}
The zero-temperature Hall response within tight-binding models of
correlated electrons is studied. Using the linear response theory and a 
linearization in the magnetic field $B$, a general relation for the
reactive (zero frequency) Hall constant $R^0_H$ in the fast
(transport) limit is derived, involving only matrix elements between
the lowest excited states at $B=0$; for noninteracting fermions, the
Boltzmann expression is reproduced. For a Fermi liquid with a well
defined Fermi surface and linear gapless excitations an analogous
expression is found more generally. In the specific case of   
quasi-one-dimensional correlated systems a relation of $R^0_H$ to the charge
stiffness $D$ is recovered. Similar analysis is performed and
discussed for $D$ and the compressibility.
\end{abstract}
\maketitle

\section{Introduction}

Properties of metals with strongly correlated electrons can be
strikingly inconsistent with the usual picture of ordinary Fermi
liquid. The most intensively studied example are the superconducting 
cuprates \cite{ong, hwan},
which behave in the normal state as hole doped magnetic (or
Mott-Hubbard) insulators. Here an evident and challenging theoretical
problem is to reconcile the Hall constant $R_H^0$, following
approximately a simple semiclassical behavior $R_H^0 \sim 1/n_h e_0$
consistent with an (semiconductor-like) interpretation of low
concentration $n_h$ of holes in an insulator, with the evidence for
large Fermi surface as it emerges e.g. from photoemission experiments.

The theoretical analysis of the Hall constant $R_H^0$ and more generally
of the dynamical Hall response $R_H(\omega)$ in systems with
correlated electrons has proved to be very difficult. Within the
general linear response theory the procedure for the calculation of
Hall response is in principle well established \cite{fuku} and
requires the introduction of a modulated magnetic field $B~{\rm
exp}(i{\bf qr})$ and consequently a modulated vector potential ${\bf
A}$ in order to formally allow a linearization of the offdiagonal
conductivity $\sigma_{\beta\alpha}$ in $B\neq 0$ and to derive an
expression for $R_H(q,\omega)$ not involving $B$. At $T=0$ the
relevant case for usual transport measurements of d.c. $R_H^0$ is the
limit: $q \to 0$ first, and then $\omega \to 0$ \cite{fuku,rojo}.  The
formulation originally designed for nearly free lectrons has been
extended to tight-binding models for strongly correlated electrons
\cite{shas}. Nevertheless there have been so far rather few results 
for strongly correlated models obtained following this framework. The
Hall mobility of a single carrier at large $T$ has been evaluated
\cite{brin}. $R_H(\omega)$ has been calculated in the high-$\omega,T$
expansion \cite{shas}, indicating on the change of sign of $R_H$ in
the vicinity of the Mott-Hubbard insulator.  One of the present authors
\cite{prel} also showed a plausible but nontrivial result that a
single carrier doped into a magnetic insulator at $T=0$ indeed follows
the semiclassical formula for $R_H^0$.

More direct numerical evaluation of $R_H^0$ within the linear response
approach (at finite $B>0$) is also quite delicate for prototype models
of correlated electrons. Namely, small-system studies give even some
controversial conclusions regarding the sign of $R_H^0$ close to the
magnetic insulator \cite{cast,assa,tsun}. It has been shown
\cite{prel1} that a better controlled approach at $T=0$ can
be obtained for a system with open boundary conditions in one
direction, i.e. on a ladder geometry, where $R_H^0$ can be expressed
in terms of derivatives of the ground state energy with respect to
external fields.

Recently, the present authors \cite{zoto} derived at $T=0$ a quite general
relation between the reactive Hall constant $R_H^0$ and the derivative
of the charge stiffness $D$ with respect to the electron density $n$, 
\begin{equation}
R_H^0= - \frac{1}{e_0 D}\frac{\partial D}{\partial n}~. \label{eq1}
\end{equation}
It should be pointed out that the derivation uses, in the direction 
transverse to the driving field, the non-standard
(slow) limit $\omega \to 0$ first, then $q \to 0$. It is a question
and also our aim to find out whether or under which conditions this 
relation applies to the more relevant fast limit: $q \to 0$, then $\omega
\to 0$. Eq.(\ref{eq1}) has some very attractive properties for the 
analysis of strongly correlated electrons: a) $D$ is at $T=0$ the
central quantity distinguishing the Mott-Hubbard insulator from a
conductor (metal), b) close to the Mott-Hubbard insulator where the
stiffness is expected to be proportional to hole doping, i.e. $D
\propto n_h=1-n$, Eq.(\ref{eq1}) directly implies the plausible
semiclassical result $R_H^0=1/e_0 n_h$, which has been hard to establish by
any other analytical method so far. 

A qualitatively similar relation to Eq.(\ref{eq1}) was also proposed in
\cite{rojo}, where $\sigma_{xy}$ was related to the variation of the
kinetic energy, i.e. $\partial
\langle T\rangle/\partial n$. Note that in a tight binding system 
$\langle T\rangle$
and $D$ are interrelated through the optical sum rule.

Our goal is to express the $T=0$ reactive Hall constant $R_H^0$ in
terms of eigenstates of correlated electrons in the absense of $B$ and
the fast limit, $q \to 0$ first and then $\omega \to 0$. The present
approach is the extention of the previous analysis for a single
carrier \cite{prel} to the general electron concentration $n$. As
first formulated by Kohn \cite{kohn}, the diagonal conductivity in a
metal is singular (reactive) at $T=0$ and low frequencies,
i.e. $\sigma_{\alpha\alpha} \propto D_{\alpha\alpha}/\omega$, defining
the charge stiffness $D_{\alpha\alpha}$. The concept of charge
stiffness has been essential in the studies of correlated systems and
is quite well established both by analytical and numerical methods in
a number of relevant models \cite{dago}. 

On the other hand, in the same system at $B\agt 0$ one also expects a
singular off-diagonal conductivity $\sigma_{\alpha\neq\beta}
\propto B \Lambda_{\alpha\beta}/\omega^2$, leading to a finite reactive
$R_H^0$. $\Lambda_{\alpha\beta}$ (independent of $B$) is thus a
central quantity of interest, playing the role of the off-diagonal
charge stiffness. We find a general expression for
$\Lambda_{\alpha\beta}$ in terms of low lying states at $B=0$, and
a more specific one for Fermi liquids with a well defined Fermi surface
and charge excitations with linear dispersion. In the latter case we discuss 
the relation of the new formulation to Eq.(\ref{eq1}) as well as to 
the standard Boltzmann theory for $R_H^0$ in metals within 
the single relaxation-time approximation \cite{jone,ong1}.

The paper is organized as follows: In Sec. II we introduce the linear
response formalism for $R_H^0$. We perform the linearization in $B$
and derive a general expression for $\Lambda$ in terms of electron
eigenstates. In Sec. III we treat in analogous way as a limit $q\to 0$
the stiffness $D^0_{\alpha\alpha}$ and the compressibility $\kappa$ in
order to demonstrate that in these quantities appear similar matrix
elements as in $\Lambda$. We also discuss the Hall conductivity
$\sigma_{yx}$ in the slow limit ($\omega \to 0$ first), thus relating
it to the generalized $D$. In Sec. IV we deal with two specific cases:
a single carrier in a Mott-Hubbard insulator and 1D systems.  Sec. V
is devoted to a derivation assuming a Fermi liquid system, obtaining
an expression analogous to the one in the relaxation-time
approximation. As the application of the analysis we discuss
noninteracting fermions, an isotropic Fermi liquid, and in particular
the quasi-1D system where we recover the relation (\ref{eq1}).

\section{Dynamical Hall response} 

For simplicity we consider only a two-dimensional system in the $x-y$
plane, with a magnetic field perpendicular to the plane, i.e. ${\bf
B}=\hat B{\bf e}_z$. To calculate the dynamical Hall constant
$R_H(\omega)$ we are looking on the response to a uniform electric
current ${\bf J}= J_x {\bf e}_x$ . We follow the standard linear
response analysis \cite{fuku}, introducing a modulated magnetic
field. Here we choose the modulation in the $y$ direction with ${\bf
q}=q {\bf e}_y$ (the final result in the limit $q \to 0$
should be independent of the ${\bf q}$ direction) requiring also a
modulated electric field with the same ${\bf q}$,
\begin{equation}
\hat B= B {\rm e}^{iqy}, \qquad
{\vec {\cal E}} ={\cal E}_y^{\bf q} {\rm e}^{iqy} {\bf e}_y~, \label{eq2}
\end{equation}
while the vector potential in the Landau gauge is
\begin{equation}
{\bf A}=A^{\bf q} {\rm e}^{iqy} {\bf e}_x~, \qquad A^{\bf q}= i B/q~.\label{eq3}
\end{equation}

The dynamic Hall response is given by \cite{shas}
\begin{equation}
R_H(\omega) =\frac{{\cal E}_y^q(\omega)}{B J_x(\omega) }=
\frac{1}{B}\frac{-\hat \sigma_{yx}(\omega)}{
\sigma_{xx}(\omega) \sigma_{yy}(\omega)}\Big|_{B\to 0,q\to 0}~,
\label{eq4}
\end{equation}
where $\hat \sigma_{\alpha\beta},\sigma_{\alpha\beta}$ denote the
conductivities at $B\ne 0$ and at $B=0$, respectively, higher
order terms in $B$ have already been neglected in Eq.(\ref{eq4}); the hat will
denote quantities in magnetic field from here on and we are interested in the 
limit $q\to 0, B\to 0$. 

Models for strongly correlated electrons are usually analysed within
the tight binding framework $H=T+H_{int}$ where the magnetic field
(flux) enters through the kinetic energy $T$ via the Peierls phase,
i.e.
\begin{equation}
T=-\sum_{(ij) s} t_{ij}[{\rm e}^{i (\theta_{ij}+\phi_{ij})}
c^{\dagger}_{js} c_{is} + H.c.]~, \label{eq5}
\end{equation}
Here it is meaningful to distinguish the phase due to a constant
magnetic field $B$ (which in principle can be large), i.e.
$\theta_{ij}=e{\bf r}_{ij}\cdot {\bf A}({\bf r}={\bf R}_{ij})$, and
the small driving (time dependent field) $\phi_{ij}(t)=e{\bf
r}_{ij}\cdot {\vec \phi}({\bf r}={\bf R}_{ij},t)$ where ${\bf
r}_{ij}={\bf r}_j - {\bf r}_i$ and ${\bf R}_{ij}=({\bf r}_i +{\bf
r}_j)/2$; the sum $(ij)$ runs over pairs of sites. The (particle)
current $J$ can be defined as,
\begin{equation}
J^{\bf p}_{\alpha}=-\frac{1}{e}\frac{\partial T }{\partial 
\phi_{\alpha}^{-{\bf p}}}=\hat j^{\bf p}_{\alpha} -
e \hat \tau^{\bf p}_{\alpha\beta} \phi^0_\beta~,  \label{eq6}
\end{equation}
where $\hat j^{\bf p}_{\alpha}$ and $\hat \tau^{\bf p}_{\alpha\beta}$
refer to the paramagnetic current and stress tensor (diamagnetic
contribution), respectively, both in the presence of finite ${\bf A}$,
\begin{eqnarray}
\hat j^{\bf p}_{\alpha} &=&\sum_{(ij) s} t_{ij} 
r^{\alpha}_{ij}{\rm e}^{i {\bf
p}\cdot {\bf R}_{ij} } (i{\rm e}^{i
\theta_{ij}} c^{\dagger}_{js} c_{is} + H.c.)~, \nonumber \\ 
\hat \tau^{\bf p}_{\alpha\beta}&=& \sum_{(ij) s}
t_{ij} r^{\alpha}_{ij} r^{\beta}_{ij } {\rm e}^{i {\bf p}\cdot {\bf
R}_{ij} } ({\rm e}^{i \theta_{ij}} c^{\dagger}_{js} c_{is} + H.c.)~.
\label{eq7}
\end{eqnarray}
The conductivity tensor at $B\ne 0$, as a linear response to 
$\phi^{0}_\beta(t)$, can be expressed as \cite{fuku,shas},
\begin{eqnarray}
\hat \sigma_{\alpha\beta}(\omega)&=&\frac{i e^2}{N
\omega} [\langle \hat \tau^{\bf q}_{\alpha\beta} \rangle
-\hat \chi_{\alpha\beta}(\omega)]~, \nonumber \\
\hat \chi_{\alpha\beta}(\omega)&=&i\int_0^\infty dt {\rm
e}^{i\omega t}  \langle [\hat j_{\alpha}^{\bf q}(t)~,
\hat j_{\beta}^0] \rangle _B~, \label{eq8}
\end{eqnarray}
where $N$ is the number of cells (volume of the unit cell is assumed
unity) and $\langle ~\rangle_B$ denote averages at  $B\ne 0$.

\subsection{Linearization in $B$}

In order to calculate $R_H(\omega)$ we have to evaluate $\hat
\sigma_{yx}$ up to the linear term in $B$. Since $\langle \hat
\tau^{\bf q}_{yx} \rangle =0$, we need
\begin{equation}
\hat \chi_{yx}(\omega) = -e A^{\bf q} K_{yx}(\omega)~. \label{eq9}
\end{equation}
In a conductor at $T=0$ the conductivities $\sigma^0_{\alpha\alpha}$
are singular at $\omega=0$ \cite{kohn} defining the charge stiffness
$D^0_{\alpha\alpha}$,
\begin{equation}  
\sigma^0_{\alpha\alpha}(\omega)= \frac{2 i  e^2}{\omega} 
D^0_{\alpha\alpha} + \sigma_{\alpha\alpha}^{reg}(\omega)~, \label{eq10}
\end{equation}
where $\sigma_{\alpha\alpha}^{reg}(\omega)$ is the regular part of the
conductivity.  In the same limit $K_{xy}(\omega)$ is also
expected to be singular leading to a finite $R_H^0=R_H(\omega \to
0)$. Hence we define the off-diagonal stiffness
\begin{equation}
\Lambda_{yx}= -\Bigl[\frac{\omega K_{yx}(\omega)}{4
Nq}\Bigr|_{q\to 0}\Bigr]_{\omega \to 0}~, \label{eq11}
\end{equation}
so that we can express the reactive Hall constant as
\begin{equation}
R_H^0=\frac{\Lambda_{yx}}{e D^0_{xx}D^0_{yy}}~. \label{eq12}
\end{equation}

Performing now the linearization in the static vector potential
$A^{\bf q}$ we take into account the analogy to Eq.(\ref{eq6}) and the
coupling to the field,
\begin{equation}
\hat j^{\bf p}_{\alpha}=j^{\bf p}_{\alpha}-e
\tau^{\bf p-q}_{\alpha x}A^{\bf q}~, \qquad
H'=- e j^{-{\bf q}}_x A^{\bf q}~, \label{eq13}
\end{equation}
where $j^{\bf p}_{\alpha}= \hat j^{\bf p}_{\alpha}(B=0)$ and
$\tau^{\bf p}_{\alpha\beta}= \hat \tau^{\bf p}_{\alpha\beta}(B=0)$.
Instead of a general formalism at $T>0$ \cite{fuku} we assume here
explicitly $T=0$ and we can express
\begin{eqnarray}
\hat \chi_{yx}&=& \langle 0_B |j^{\bf q}_y \frac{1}{\omega +\hat E_0 -
\hat H}j^{0}_x| 0_B \rangle \nonumber \\
&-&\langle 0_B |j^{0}_x
\frac{1}{\omega -\hat E_0 + \hat H}j^{\bf q}_y| 0_B \rangle=
\chi_{yx}^I+\chi_{yx}^{II}~, \label{eq14}
\end{eqnarray}
where $| 0_B \rangle,\hat E_0,\hat H$ refer to $B\neq 0$. We
consider only linear terms in $A^{\bf q}$ therefore $\hat E_0=E_0$.
Taking into account that
\begin{eqnarray}
&&|0_B \rangle\sim |0 \rangle - eA^{\bf q} \frac{1}{E_0-H}
j^{-\bf q}_x| 0 \rangle~, \nonumber\\ 
&&\frac{1}{X-Y} \sim \frac{1}{X}+ \frac{1}{X}Y \frac{1}{X}~, \label{eq15}
\end{eqnarray}
we obtain
\begin{eqnarray}
K_{yx}^I&=& \langle 0|j^{\bf q}_y \frac{1}{\omega- (H-E_0)}
\tau^{-{\bf q}}_{xx}|0 \rangle \nonumber \\ 
&+& \langle 0|\tau^{0}_{yx}
\frac{1}{\omega - (H-E_0)} j^{0}_x|0 \rangle  \nonumber \\ 
&+&\langle 0|j^{\bf q}_y \frac{1}{\omega - (H-E_0)}j^{0}_x 
\frac{1}{E_0-H}  j^{-\bf q}_x| 0\rangle  \label{eq16} \\
&+& \langle 0|j^{-{\bf q}}_x \frac{1}{E_0-H}
j^{\bf q}_y \frac{1}{\omega - (H-E_0)}j^{0}_x|0 \rangle  
\nonumber \\ 
&+&\langle 0|j^{\bf q}_y \frac{1}{\omega - (H-E_0)}
j^{-\bf q}_x \frac{1}{\omega - (H-E_0)}j^{0}_x|0 \rangle~, \nonumber
\end{eqnarray}
with an analogous expression holding for $K_{yx}^{II}$. 

\subsection{Fast limit}

It is helpful to express $K_{yx}(\omega)$ in terms of
eigenstates. Assuming that the ground state $|0\rangle$ has momentum
${\bf Q}$ it is convenient to separate excited states into sectors:
$|m\rangle$ with momentum ${\bf Q}$, $|\tilde m\rangle$ with ${\bf
Q}-{\bf q}$, and $|\bar m\rangle$ with ${\bf Q}+{\bf q}$.  Denoting
$\epsilon=E-E_0$ and using
\begin{equation}
\frac{1}{(\omega-\epsilon_{\tilde m})(\omega-\epsilon_l)}=
\frac{1}{\epsilon_{\tilde m}-\epsilon_l}\Bigl(
\frac{1}{\omega-\epsilon_{\tilde m}}-
\frac{1}{\omega-\epsilon_l} \Bigr)~, \label{eq17}
\end{equation}
we can write \cite{prel}
\begin{eqnarray}
K_{yx}(\omega)&=&\sum_{\tilde m}\Bigl[\frac{\gamma_{\tilde
m}} {\omega - \epsilon_{\tilde m}} +\frac{\tilde \gamma_{\tilde m}}
{\omega + \epsilon_{\tilde m}} \Bigr] \nonumber\\
&+& \sum_{m>0}\Bigl[\frac{\delta_m }{\omega - \epsilon_m} +
\frac {\tilde \delta_m }{\omega + \epsilon_m} \Bigr]~, \label{eq18}
\end{eqnarray}
where 
\begin{eqnarray}
\gamma_{\tilde m}&=&2(j^{\bf q}_y)_{0\tilde m}(d^{-{\bf
q}}_{xx})_{\tilde m 0}, \nonumber \\ 2(d^{-{\bf q}}_{xx})_{\tilde m
0}&=& (\tau^{-{\bf q}}_{xx})_{\tilde m 0}- \sum_{l>0} \frac{(j^{-{\bf
q}}_x)_{\tilde m l}(j^0_x)_{l0}}{\epsilon_l - \epsilon_{\tilde m} } 
\nonumber \\ &-& \sum_{\tilde l} \frac{(j^0_x)_{\tilde m \tilde l}
(j^{-{\bf q}}_x)_{\tilde l 0} } {\epsilon_{\tilde l} }~, \label{eq19}
\end{eqnarray}
and
\begin{eqnarray}
\delta_m&=&2(d^0_{yx})_{0 m}(j^0_x)_{m 0},\nonumber \\
2(d^0_{yx})_{0 m}&=& (\tau^0_{yx})_{0m} - \sum_{\tilde l}
\frac{(j^{\bf q}_y)_{0\tilde l} (j^{-{\bf q}}_x)_{\tilde l m}}
{\epsilon_{\tilde l} - \epsilon_m} \nonumber \\ 
&-&\sum_{\bar l}\frac{(j^{-{\bf q}}_x)_{0 \bar l} (j^{\bf q}_y)_{\bar l m}}
{\epsilon_{\bar l}}~, \label{eq20}
\end{eqnarray}
and analogous expressions for $\tilde \gamma_{\tilde m}$ and $\tilde
\delta_m$. Since we are interested in the limit $q \to 0$ it follows
anyway from $\tilde \sigma_{\alpha\beta}(-\omega)= \tilde
\sigma^*_{\alpha\beta}(\omega)$ that $\tilde \gamma_{\tilde
m}=\gamma_{\tilde m} = {\rm real}$ and $\tilde \delta_m = \delta_m =
{\rm real}$.

The fast limit corresponds taking first $q\to 0$ and then $\omega\to
0$. In Eq.(\ref{eq18}) the singular part $\propto 1/\omega$ in this case
emerges from the class of excited states $\tilde M$ which exhibit
$\epsilon_{\tilde m} \to 0$ in the $q \to 0$ limit. On the other hand
$\delta_m, \tilde \delta_m$ terms are not contributing since
$\epsilon_m>0$ do not depend on $q$ and remain finite in the limit
$q\to 0$. $\Lambda_{yx}$ can therefore be expressed as
\begin{eqnarray}
\Lambda_{yx} &=& \lim_{q\to 0} \frac{1}{2Nq}\sum_{\tilde m \in
\tilde M} \gamma_{\tilde m} = \nonumber \\ 
&=& \lim_{q\to 0} \frac{1}{Nq}\sum_{\tilde m \in \tilde M} (j^{\bf
q}_y)_{0\tilde m} (d^{-{\bf q}}_{xx})_{\tilde m 0}~. \label{eq21}
\end{eqnarray}

We note that $(d^{-{\bf q}}_{\alpha\alpha})_{\tilde m n}$ can also be 
represented as the matrix element of the stiffness operator $
d_{\alpha\alpha}^{-{\bf q}}$,
\begin{eqnarray}
&&(d^{-{\bf q}}_{\alpha\alpha})_{\tilde m n}= \langle \tilde m|
d^{-{\bf q}}_{\alpha\alpha} |n \rangle =
\frac{1}{2}\langle \tilde m| [\tau^{-{\bf q}}_{\alpha\alpha} \nonumber \\
&& -j^{-{\bf q}}_\alpha \frac{1}{H-E_{\tilde m}} j^0_\alpha
- j^{0}_\alpha \frac{1}{H-E_n} j^{-{\bf q}}_\alpha] |n \rangle~,
\label{eq22}
\end{eqnarray}
provided that $(j^0_\alpha)_{nn}=0$. It is quite evident that the
operator $\hat d^{-{\bf q}}$ is closely related to the charge
stiffness since it follows from Eq.(\ref{eq8}),(\ref{eq10}) that
$D^0_{\alpha\alpha}=\langle 0|d^0_{\alpha\alpha} |0 \rangle/N$.

Moreover, there are other more compact representations of $(d^{-{\bf
q}}_{xx})_{\tilde m 0}$.  We first note that it just the matrix
element at $B>0$,
\begin{equation}
(d^{-{\bf q}}_{xx})_{\tilde m 0}= \frac{iq}{2}\frac{\partial}{\partial B}
\langle m_B|\hat j^0_x |0_B \rangle~, \label{eq22a}
\end{equation}
as follows directly from Eq.(\ref{eq14}) by using the representation
of $|m_B\rangle$ states and extracting the singular term at $q\to 0$, 
\begin{eqnarray}
\chi_{yx}^I&=& \sum_{m_B} \langle 0_B|\hat j^{\bf q}_y|m_B\rangle
\frac{1}{\omega+\hat E_0-\hat E_m}\langle m_B|\hat j^0_x|0_B
\rangle \nonumber \\
&\to& \frac{1}{\omega} \sum_{\tilde m \in \tilde M} \langle 0|j^{\bf
q}_y| \tilde m\rangle \langle m_B|\hat j^0_x|0_B \rangle~, \label{eq22b}
\end{eqnarray} 
The latter emerges from $m_B$ corresponding to $\tilde m$ and the only
term linear in $B$ involves Eq.(\ref{eq22a}).

$(d^{-{\bf q}}_{xx})_{\tilde m 0}$ can be as well interpreted as a
derivative of the current matrix elements with respect to a uniform
vector potential (fictitious flux) $\vec \theta$
\cite{kohn}, coupled to the current as $H''= {\vec \theta}\cdot {\bf
j}$. By taking a derivative with respect to $\theta_x$, we obtain 
\begin{eqnarray}
&&\frac{\partial}{\partial \theta_x}\langle \tilde m|j_x^{-{\bf
q}}|0\rangle= \langle \tilde m|\Bigl[\tau^{-{\bf q} }_{xx}\nonumber
\\ && -\sum_{l>0}
\frac{j_x^{-{\bf q}}|l\rangle \langle l|
j_x^0}{\epsilon_l}-\sum_{\tilde l \neq \tilde m} \frac{j_x^0| \tilde
l\rangle \langle \tilde l| j_x^{-{\bf q}}}{\epsilon_{\tilde
m}-\epsilon_{\tilde l}} \Bigr] |0\rangle~.  \label{eq23}
\end{eqnarray}
In the limit $q\to 0$ we can replace  $\epsilon_l\to
\epsilon_l-\epsilon_{\tilde m}$ for $l>0$, and $\epsilon_{\tilde
l}-\epsilon_{\tilde m} \to \epsilon_{\tilde l}$ for $\tilde l \neq
\tilde m$. Since $(j_x^0)_{00}=0$ we can identify Eq.(\ref{eq23}) with
Eq.(\ref{eq19}) except the term $\tilde l=\tilde m$,
\begin{equation}
2 (d^{-{\bf q}}_{xx})_{\tilde m 0} = \frac{\partial}{\partial
\theta_x}\langle \tilde m|j_x^{-{\bf q}}|0\rangle -
\frac{(j^0_x)_{\tilde m \tilde m} (j^{-{\bf q}}_x)_{\tilde m 0} }
{\epsilon_{\tilde m} }~.  \label{eq24}
\end{equation}

For a general correlated system we have thus expressed $\Lambda_{xy}$
and consequently $R^0_H$ in terms of matrix elements involving only
lowest excited states $\tilde m \in \tilde M$ (in the absense of $B$)
and their derivatives to a homogenenous flux. In addition we note that
the second term in Eq.(\ref{eq24}) does not contribute to
$\Lambda_{xy}$ in several nontrivial cases as shown later: a) for a
single carrier in a correlated system, b) for noninteracting fermions
on a bipartite lattice with only nearest neighbor hopping.

\section{Charge stiffness and compressibility}

Before proceeding with the approximations to $\Lambda_{xy}$ let us
stress that also other quantities like the charge stiffness and the
compressibility can be expressed solely in terms of the same excited
states $\tilde m \in \tilde M$.

We first consider $\sigma^0_{\alpha\alpha}$ for $B=0$.
$D^0_{\alpha\alpha}$ can be expressed as
\begin{equation}
D^0_{\alpha\alpha}= \frac{1}{N}\Bigl[ \frac{\langle
\tau^0_{\alpha\alpha}\rangle}{2} -
\sum_{m>0} \frac{|(j^0_{\alpha})_{0m}|^2}{\epsilon_m} \Bigr]
=\frac{1}{N} \frac{\partial^2 E_0(\theta_\alpha)}{\partial^2 \theta_\alpha}~,
\label{eq25}
\end{equation}
hence $D^0_{\alpha\alpha}$ is evaluated from the ground state energy
$E^0(\theta_\alpha)$, which is the usual procedure.

Alternatively we can consider $\sigma^0_{\alpha\alpha}$ as the limit
$q\to 0$ of $\sigma^{\bf q}_{\alpha\alpha}$ (direction of ${\bf q}$
here is arbitrary),
\begin{eqnarray}
\sigma^{\bf q}_{\alpha\alpha}(\omega)&=&\frac{i e^2}{N \omega} [\langle
\tau^0_{\alpha\alpha} \rangle -
\chi^{\bf q}_{\alpha\alpha}(\omega)]~, \nonumber \\
\chi^{\bf q}_{\alpha\alpha}(\omega)&=&i\int_0^\infty dt {\rm e}^{i\omega
t} \langle [j_{\alpha}^{\bf q}(t), j_{\alpha}^{-{\bf q}}]
\rangle~,  \label{eq26}
\end{eqnarray}
In the case of $q\neq 0$ it follows $\langle \tau^0_{\alpha\alpha}
\rangle=\chi^{\bf q}_{\alpha\alpha}(0)$ (optical sum rule), so there
is no (strictly) reactive term which would correpond to the
singularity in Eq.(\ref{eq10}). As in Sec. II we can then represent
$\sigma^{\bf q}_{\alpha\alpha}$ in terms of eigenstates,
\begin{equation}
\sigma^{\bf q}_{\alpha\alpha}(\omega)= \frac{i e^2}{N} \sum_{\tilde
m} \frac{|(j_\alpha^{\bf q})_{0\tilde m}| ^2}{\epsilon_{\tilde m}} \Bigl
[\frac{1}{\omega - \epsilon_{\tilde m}} +\frac{1}{\omega +
\epsilon_{\tilde m}} \Bigr]~.  \label{eq27}
\end{equation}
We can nevertheless extract the reactive part as the singular part
which behaves as $1/\omega$ for $q\to 0$. This again emerges from
states $\tilde m \in \tilde M$ and consistent with Eq.(\ref{eq10}) we
get  
\begin{equation}
D^0_{\alpha\alpha} = \lim_{q\to 0} \frac{1}{N}\sum_{\tilde m \in
\tilde M} \frac{|(j_\alpha^{\bf q})_{0\tilde m}| ^2}{\epsilon_{\tilde
m}}~.  \label{eq28}
\end{equation}
Eqs. (\ref{eq25}) and (\ref{eq28}) define two alternative approaches
to evaluate $D^0_{\alpha\alpha}$, where the first is the standard
one. The equivalence of both is expected to give more insight into the
excited states and matrix elements $(j_\alpha^{\bf q})_{0\tilde
m}$. On the other hand we note that $D^0_{\alpha\alpha}$,
Eq.(\ref{eq28}), contains the same matrix elements as $\Lambda_{yx}$
indicating that both quantities as well as $R_H^0$ are related.

Closely related is also the generalized compressibility $\kappa^{\bf
q}$. Let us consider a perturbation induced by the modulated chemical
potential so that we deal with a hamiltonian $H^\mu=H+ \mu^{\bf q}
\rho^{\bf q}$. Then we can express in analogy to $D^{\bf
q}_{\alpha\alpha}$,
\begin{equation}
\kappa^{\bf q}=\frac{1}{N}\frac{\partial \langle \rho^{\bf q}\rangle }
{\partial \mu^{\bf q}}=\frac {2}{N} \sum_{\tilde m} 
\frac{|(\rho^{\bf q})_{0\tilde m}|^2} {\epsilon_{\tilde m}}~.  \label{eq29}
\end{equation}

We also note the relation following from the conservation law for
$q \to 0$,
\begin{eqnarray}
&&i[H, \rho^{\bf q}]+i {\bf q}\cdot {\bf j}^{\bf q}=0 \nonumber \\
&&\longrightarrow {\bf q}\cdot ({\bf j}^{\bf q})_{l\tilde m} =
(\epsilon_{\tilde m}- \epsilon_l) (\rho^{\bf q})_{l\tilde m}~,
\label{eq30}
\end{eqnarray}
so $({\bf j}^{\bf q})_{0\tilde m}$ and $(\rho^{\bf q})_{0\tilde m}$
are evidently related.  From Eq.(\ref{eq30}) it follows that
$(\rho^{\bf q})_{0\tilde m} \propto q$ for $\tilde m \notin \tilde M$, so
only states $\tilde m \in \tilde M$ contribute in the limit $q \to 0$,
\begin{equation}
\kappa^0=\lim_{q\to 0} \kappa^{\bf q} =\lim_{q\to 0} \frac{2}{N}
\sum_{\tilde m \in \tilde M} \frac{|(\rho^{\bf q})_{0\tilde m}|^2}
{\epsilon_{\tilde m}}~.  \label{eq31}
\end{equation}

In order to find a closer relation between $\Lambda_{yx}$ and
the stiffness $D_{\alpha\alpha}$ as e.g. manifested in Eq.(\ref{eq1}) let
us consider now the stiffness ${\cal D}_{xx}^{-\bf q}$ in a perturbed
ground state \cite{zoto},
\begin{eqnarray}
2{\cal D}_{xx}^{-\bf q}=&&\frac{1}{N} \langle 0_\mu|\Bigl[ \tau^{-{\bf q}
}_{xx}- j_x^0 \frac{1}{H^\mu-E^\mu_0} j_x^{-\bf q} \nonumber \\
&&- j_x^{-{\bf q}} \frac{1}{H^ \mu-E^\mu_0} j_x^{0} \Bigr] |0_\mu\rangle~,
\label{eq32}
\end{eqnarray}
We can again evaluate Eq.(\ref{eq32}) within the first order of the
perturbation theory in $H'=\mu^{\bf q} \rho^{\bf q}$
\begin{equation}
|0_\mu\rangle= |0\rangle + \sum_{\tilde m} |\tilde m\rangle
\frac{(\rho^{\bf q})_{\tilde m 0} \mu^{\bf q}}{-\epsilon_{\tilde
m}} \label{eq33}
\end{equation}
and recognize the correspondence with $\tilde \sigma_{yx}$ as
explicitly expressed in Eqs.(\ref{eq8}),(\ref{eq9}),(\ref{eq17}) -
(\ref{eq20}). In fact it has been already shown \cite{zoto} that 
\begin{equation}
\frac{\partial {\cal D}_{xx}^{-{\bf q}}}{\partial \mu^{\bf q}}= 
\frac{q^2}{e^3B} \tilde \sigma_{yx}(0)~, \label{eq34}
\end{equation}
Using Eqs.(\ref{eq8}),(\ref{eq18}) and decomposing
$1/[\omega(\omega-\epsilon_m)]$ as in Eq.(\ref{eq17}), we can rewrite
\begin{eqnarray}
\tilde\sigma_{yx}(\omega) = &&\frac{e^3B}{qN}\Bigl\{
\sum_{\tilde m}\frac{1}{\epsilon_{\tilde m}}\Bigl[
\frac{\gamma_{\tilde m}} {\omega - \epsilon_{\tilde m}} 
-\frac{\tilde \gamma_{\tilde m}}
{\omega + \epsilon_{\tilde m}} \Bigr] \nonumber \\
&&+ \sum_{m>0}\frac{1}{\epsilon_m}\Bigl[\frac{\delta_m }
{\omega - \epsilon_m} + \frac {\tilde \delta_m }
{\omega + \epsilon_m} \Bigr] \Bigr\}~,  \label{eq35}
\end{eqnarray}
taking into account that at $q>0$ there is no singularity strictly at
$\omega=0$, hence terms $1/\omega$ should cancel. For $\omega=0$ we get
\begin{equation}
\tilde\sigma_{yx}(0) = \frac{2 e^3B}{qN} \Bigl[ \sum_{\tilde m}
\frac{\gamma_{\tilde m}}{\epsilon_{\tilde m}^2} + 
\sum_{m>0}\frac{\delta_m}{\epsilon_m^2} \Bigr]~.
\label{eq36}
\end{equation}
It seems plausible that in the limit $q \to 0$ in Eq.(\ref{eq36}) only
states with $\epsilon_{\tilde m} \propto q$ contribute, i.e. $\tilde m
\in \tilde M$, so that
\begin{equation}
\frac{\partial {\cal D}_{xx}^{-{\bf q}}}{\partial \mu^{\bf q}}=
\frac{2 q}{N} \sum_{\tilde m \in \tilde M}\frac{\gamma_{\tilde m}}
{\epsilon_{\tilde m}^2}~. \label{eq37}
\end{equation}
The relation to $\Lambda_{yx}$ in Eq.(\ref{eq21}) is evident and will
be exploited later on.

\section{Specific cases}

\subsection{Single charge carrier}

A nontrivial example of the above formalism is a single charge carrier
i.e. a hole or an electron doped into a Mott-Hubbard insulator
\cite{prel}. We have to assume only that the carrier behaves as a
quasiparticle. I.e., excited states have a well defined effective mass,
\begin{equation}
\epsilon_{\tilde m}= N D^0_{yy} q^2~. \label{eq52}
\end{equation}
For a nondegenerate ground state $|0\rangle$ there is only one
relevant excited state $|\tilde m\rangle=|\tilde 0 \rangle$. So it
follows from Eqs.(\ref{eq28}),(\ref{eq52}),(\ref{eq21}) that
\begin{equation}
D^0_{yy}=\frac{(j^{\bf q}_y)_{0\tilde 0}^2}{N \epsilon_{\tilde 0}}
\quad \to \quad |(j^{\bf q}_y)_{0\tilde 0}|= N D^0_{yy} q~, \label{eq53}
\end{equation}
and
\begin{equation}
\Lambda_{yx}=\frac{1}{N q}(j^{\bf q}_y)_{0\tilde 0}(d^{-{\bf q}}_{xx})_{\tilde
0 0} = \pm D^0_{yy} D^0_{xx}~. \label{eq54}
\end{equation}
The semiclassical result follows finally from Eq.(\ref{eq12}),
\begin{equation}
R_H^0 = \pm \frac{N}{e_0}~,\qquad {\rm sgn}(R^0_H)=
-{\rm sgn} (j^q_y)_{0\tilde 0}~, \label{eq54a}
\end{equation}
by inserting $e=-e_0$. There remains to determine the sign of
$R_H^0$, which should be plausibly positive for a hole-doped insulator
although this is not trivial to show analytically \cite{prel}.  For a single
carrier we also get for $q\to 0$ $(\rho^q)_{0\tilde0}=1$ which is an 
alternative requirement for a well defined quasiparticle. We note also
that the second term in Eq.(\ref{eq24}) does not contribute since
$(j_x^0)_{\tilde 0 \tilde 0}=0$ if $x$ and $y$ are symmetry directions of
the $D$ tensor.

\subsection{1D systems}

Naturally one cannot discuss Hall effect and $R_H^0$ in a strictly 1D
system but it is instructive to consider relations which follow from
our analysis for $D^0$ and $\kappa^0$. We assume here that the
correlated electron system behaves as a Luttinger liquid with 
gapless charge excitations characterized by a linear dispersion
for $q \to 0$,
\begin{equation}
\epsilon_{\tilde m}=E_{\tilde m}(q) -E_0 \sim v_c q~. \label{eq54b}
\end{equation}
The counting of states
$|\tilde m\rangle$ is then as for electron-hole excitations in the
normal Fermi liquid.

Assuming that Eqs.(\ref{eq28}),(\ref{eq31}) behave regularly as
$q\to 0$, we can replace $(j^{\bf q})_{0\tilde m} \to j_c$ and
$(\rho^{\bf q})_{0\tilde m} \to r_c$. From Eq.(\ref{eq30}) it follows
also that $j_c=v_c r_c$ and we get (taking into account also the spin
degeneracy),
\begin{equation}
D^0= \frac{j_c^2}{\pi v_c}= \frac{r_c^2 v_c}{\pi}~, \qquad
\kappa^0= \frac{2 r_c^2}{\pi v_c}~. \label{eq55}
\end{equation}
These expressions are in agreement with the phenomenology of the
Luttinger liquids \cite{hald} where we can identify the
renormalization factor $r_c$ with the density exponent
$K_\rho=r_c^2$. Although there is another gapless branch of spin
excitations, we note that this does not enter the quantities as the
charge stiffness $D^0$ and the charge compressibility $\kappa^0$. Our
analysis is at $T=0$, it is easy to argue that for low $T$
specific-heat coefficient we get $C_V \propto 1/v_c$.

\section{Fermi liquid}

Let us now consider as an illustration of the above formalism a Fermi
system characterized by gapless
charge excitations with a linear dispersion (for $q\to 0$),
\begin{equation}
\epsilon_{\tilde m}=E_{\tilde m}({\bf q}) -E_0= 
{\bf v}({\bf k})\cdot {\bf q}~, \label{eq38}
\end{equation}
corresponding to electron-hole excitations and a Fermi
surface ${\bf k}_F$. The states $|\tilde m\rangle$ are then determined
as electron-hole excitations in the normal Fermi liquid so at
given ${\bf q}$ they are characterized by ${\bf k} \in {\rm FS}_{\bf
q}$.  For the general direction ${\bf e}_\beta= {\bf q}/q$ the states
are given by ${\bf k}={\bf k}_F + \tilde k {\bf e}_\beta$ and
$-q<\tilde k<0$, where the latter condition is satisfied only along
half of the Fermi surface.

Assuming the fermionic character of such excitations (with spin), we
can write the sums over excited states explicitly for the
thermodynamic limit taking into account spin degeneracy and $q
\to 0$,
\begin{equation}
\frac{1}{N}\sum_{\tilde m}=\frac{1}{N}\sum_{{\bf k} \in {\rm FS}_q}= 
\frac{2 q}{(2\pi)^2}\oint_{k_\beta<0} \frac
{d k_F} {v({\bf k})} v_\beta({\bf k})~, \label{eq39}
\end{equation}
again restricting our analysis to 2D systems.

Here we note that in general the operators $j_{\alpha}^{\bf q}, 
\tau_{\alpha\beta}^{\bf q}, \rho^{\bf q}$ (at $B=0$) can be
represented as
\begin{eqnarray}
j_{\alpha}^{\bf q}&=&\sum_{{\bf k},s} v^\alpha_{\bf k} c^\dagger_
{{\bf k}+{\bf q}/2,s} c_{{\bf k}-{\bf q}/2,s}~, \nonumber \\
\tau_{\alpha\beta}^{\bf q}&=&\sum_{{\bf k},s} \tau^{\alpha\beta}_{\bf k}
c^\dagger_ {{\bf k}+{\bf q}/2,s} c_{{\bf k}-{\bf q}/2,s}~,
\label{eq48} \\
\rho^{\bf q}&=&\sum_{{\bf k},s} c^\dagger_{{\bf k}+{\bf q}/2,s} 
c_{{\bf k}-{\bf q}/2,s}~. \nonumber
\end{eqnarray}
where ${\bf v}_{\bf k}=\partial \epsilon_{\bf k}/\partial {\bf k}$ and
$\tau_{\bf k}=\partial^2 \epsilon_{\bf k}/ \partial {\bf k} \partial
{\bf k}$. 

In the following we can consider as a test noninteracting electrons with
dispersion $\epsilon_{\bf k}$, where the relevant excited states are
\begin{equation}
|\tilde m\rangle=c^\dagger_{{\bf k}-{\bf q}/2,s}
c_{{\bf k}+{\bf q}/2,s}|0\rangle~. \label{eq49}
\end{equation}

Let us first treat $D^{0}_{\alpha\alpha}$, Eq.(\ref{eq28}),
\begin{eqnarray}
D^{0}_{\alpha\alpha}&=& \frac{2 q}{(2\pi)^2}\oint_{k_\beta<0}
\frac {d k_F} {v({\bf k})} v_\beta({\bf k}) \frac{|(j^{\bf
q}_\alpha)_{0\tilde m}|^2}{\epsilon_{\tilde m}} = \nonumber \\ &=&
\frac{1}{(2\pi)^2}\oint \frac {d k_F} {v({\bf k})} |(j^{\bf
q}_\alpha)_{0\tilde m}|^2~. \label{eq40}
\end{eqnarray}
For $q\to 0$ the result must be independent of ${\bf q}$ so it is
plausible that
\begin{equation}
(j^{\bf q}_\alpha)_{0\tilde m} \to j_\alpha({\bf k}) \label{eq41}
\end{equation}
depends only on ${\bf k} \in {\bf k}_F$, and
\begin{equation}
D^0_{\alpha\alpha}=\frac{1}{(2\pi)^2}\oint \frac {dk_F} {v({\bf k})}
|j_\alpha({\bf k})|^2~. \label{eq42}
\end{equation}
For noninteracting fermions the expression (\ref{eq42}) is
straightforward since we know from Eqs.(\ref{eq48}),(\ref{eq49}) that
$j_{\alpha}({\bf k}) =v_{\alpha}({\bf k})=v^{\alpha}_{\bf k}$.

In the same way we can also argue that $(\rho^{\bf q})_{0\tilde m} \to
r({\bf k})$, which can be concluded from Eq.(\ref{eq29}),
\begin{eqnarray}
&&\kappa^0=\frac{q}{\pi^2}\oint_{k_\beta<0} \frac {d k_F} {v({\bf
k})} v_\beta({\bf k}) \frac{|(\rho^{\bf q})_{0\tilde
m}|^2}{\epsilon_{\tilde m}}= \nonumber \\ =&&\frac{1}{2\pi^2}\oint
\frac {d k_F} {v({\bf k})} |(\rho^{\bf q})_{0\tilde m}|^2 =
\frac{1}{2\pi^2}\oint \frac {d k_F} {v({\bf k})} r^2({\bf k})~.
\label{eq43}
\end{eqnarray}
Furthermore for noninteracting fermions we get $(\rho^{\bf q})_{0\tilde
m} = r({\bf k})=1$.

Let us turn to the discussion of $\Lambda_{yx}$. Matrix elements
$(d_{xx}^{-{\bf q}})_{\tilde m 0}$ within a Fermi liquid are (for
chosen ${\bf q}$ direction) expected to depend only on ${\bf k}$ along
the Fermi surface, so we can replace $(d_{xx}^{-{\bf q}})_{\tilde m 0}
\to d_{xx}({\bf k})$ and
\begin{equation}
\Lambda_{yx} = \frac{1}{(2\pi)^2}\oint \frac {dk_F} {v({\bf k})}
 j_y({\bf k}) v_y({\bf k}) d_{xx}({\bf k})~. \label{eq43a}
\end{equation}
Taking into account Eq.(\ref{eq24}) and that the role of
${\vec\theta}$ is to shift ${\bf k}$ we can also relate,
\begin{equation}
\frac{\partial}{\partial \theta_x} \langle \tilde m| j^{-{\bf
q}}_x|0\rangle \to \frac{\partial}{\partial k_x} j_x({\bf k})~. 
\label{eq44}
\end{equation}
Eq.(\ref{eq21}) can therefore be written as
\begin{eqnarray}
\Lambda_{yx} = &&\frac{1}{8\pi^2}\oint \frac {dk_F} {v({\bf k})}
 j_y({\bf k}) \nonumber \\
&& \Bigl[v_y({\bf k})\frac{\partial}{\partial k_x} j_x({\bf k})
-\frac{1}{q}(j^0_x)_{\tilde m\tilde m}
(j^{-{\bf q}}_x)_{\tilde m 0} \Bigr]~. \label{eq45}
\end{eqnarray}
At this stage we are unable to put also the second term in analogous
form as the first one. Still the expression is very similar to the
symmetric one,
\begin{eqnarray}
\bar \Lambda_{yx} &=&\frac{1}{8\pi^2}\oint \frac {dk_F} {v({\bf
k})} j_y({\bf k}) \nonumber\\
&&\Bigl[v_y({\bf k})
\frac{\partial}{\partial k_x} j_x({\bf k}) - v_x({\bf k})
\frac{\partial}{\partial k_y} j_x({\bf k})\Bigr]\nonumber \\
&=&\frac{1}{8\pi^2}\oint \frac {dk_F} {v({\bf
k})} j_y({\bf k}) [{\bf v}({\bf k})\times {\bf e}_B] \cdot \nabla j_x({\bf k})~,
\label{eq46}
\end{eqnarray}
which is formally equivalent to the Boltzmann expression within the
relaxation-time approximation \cite{jone,ong1}. A clear advantage of
the symmetric expression is that the required $xy$ symmetry
$\Lambda_{yx}=-\Lambda_{xy}$ is evident, since Eq.(\ref{eq46}) can be
represented as
\begin{equation}
\bar \Lambda_{yx} = \frac{1}{16 \pi^2} \oint
[d{\bf j}({\bf k})\times {\bf j}({\bf k})]|_z= 
\pm \frac{S_j}{8 \pi^2}~,
\label{eq47}
\end{equation}
where $S_j$ is the area spanned by the vector ${\bf j}({\bf k}_F)$.

Again testing with the case of noninteracting fermions we note that in
Eq.(\ref{eq24}) the second term is 
\begin{equation}
\frac{1}{q}(j^{-{\bf q}}_x)_{\tilde m 0}(j^0_x)_{\tilde m\tilde m} 
\to v_x({\bf  k}) \frac{\partial}{\partial k_y} v_x({\bf k})~, 
\label{eq51}
\end{equation}
therefore we reproduce the usual semiclassical expression \cite{ong1}
for $\bar \Lambda_{yx}$, Eqs.(\ref{eq46}),(\ref{eq47}), up to a
constant (relaxation time) which anyhow cancels out in $R_H^0$,
Eq.(\ref{eq12}). It should be also reminded that for noninteracting
fermions on a bipartite lattices with nearest neighbor hopping the
term (\ref{eq51}) vanishes since we have $v_\alpha(k_\alpha)$.

\subsection{Quasi-1D systems}

Let us assume a very anisotropic Fermi liquid with a
dispersion large only in the $x$ direction and consequently also
a nearly flat Fermi surface with $|{\bf k}_F| \sim k_0$. It is plausible
that for large anisotropy Eq.(\ref{eq43a}) can be decoupled as
\begin{equation}
\Lambda_{yx} \sim d_{xx}({\bf k}_F) \frac{1}{(2 \pi)^2} 
\oint \frac{dk_F}{v({\bf k})} v_y({\bf k}) j_y({\bf k})~.  
\label{eq56}
\end{equation}
It follows also that
\begin{equation}
D^0_{yy} =  \frac{1}{(2\pi)^2} \oint \frac{dk_F}{v({\bf k})} 
j^2_y({\bf k})~, \qquad D^0_{xx} \sim D~. \label{eq57}
\end{equation}
Now we can use relations (\ref{eq34}),(\ref{eq37}) for a Fermi liquid
in the limit $q \to 0$,
\begin{equation}
\frac{\partial {\cal D}_{xx}^{-{\bf q}}}{\partial \mu^{\bf q}}=
\frac{1}{2 \pi^2} \oint \frac{dk_F}{v({\bf k})} r({\bf
k}){d_{xx}({\bf k})}~. \label{eq58}
\end{equation}
For a nearly flat Fermi surface we can replace $d_{xx}({\bf k}) \sim
d_{xx}({\bf k}_F)$ and we get  
\begin{equation}
\frac{\partial {\cal D}_{xx}^{-{\bf q}}}{\partial \mu^{\bf q}}
\longrightarrow \frac{\partial D}{\partial \mu} =  
d_{xx}({\bf k}_F) \frac{1}{2 \pi^2} \oint \frac{dk_F}{v({\bf k})}
r({\bf k})~,\label{eq59}
\end{equation}
where we have also assumed that taking the derivative $\partial
D/\partial \mu$ is regular for $q\to 0$. So finally we can express
$R_H^0$ as ($e=-e_0$),
\begin{eqnarray}
&&R_H^0= -\frac{A}{e_0 D} \frac{\partial D}{\partial n}, \nonumber \\
A=&&\oint\frac{dk_F}{v({\bf k})} r^2({\bf k})
\oint\frac{dk_F}{v({\bf k})} v_y({\bf k}) j_y({\bf k}) / \nonumber\\ 
 &&\oint\frac{dk_F}{v({\bf k})} r({\bf k}) \oint\frac{dk_F}{v({\bf k})} 
j_y^2({\bf k})~. \label{eq60}
\end{eqnarray}
For a quasi-1D system we expect that $r({\bf k}_F) \sim r_F$, $v({\bf
k}_F) \sim v_F$ and therefore ${\bf j}({\bf k}_F) =r_F {\bf v}({\bf
k}_F)$ and consequently $A \sim 1$. Hence we have reproduced in this
case the desired expression (\ref{eq1}).

\subsection{Isotropic Fermi liquid}

Although the tight binding model, as introduced initially in
Eq.(\ref{eq5}), does not lead to an isotropic Fermi surface, an isotropic
Fermi liquid can still be of interest for illustration and can also
emerge in specific cases.  We assume here that ${\bf j}({\bf k})=j(k)
{\bf e}_{\bf k}$ and ${\bf v}({\bf k})=v(k) {\bf e}_{\bf k}$, so that
Eqs.(\ref{eq42}),(\ref{eq47}) lead to 
\begin{equation}
D^0_{\alpha\alpha}=\frac{k_F j_F^2}{4\pi v_F}~,\qquad
\bar \Lambda_{yx} = \frac{j_F^2}{4 \pi}~.
\label{eq61}
\end{equation}
and
\begin{equation}
R_H^0= - \frac{4\pi v_F^2}{e_0 k_F^2 j^2_F}=
-\frac{1}{e_0 n_F r_F^2}~, \label{eq62}
\end{equation}
where $n_F=k_F^2/2\pi$ is the effective density of electrons as
determined by the volume of the Fermi surface.

\section{Discussion}

The theory of the Hall constant in systems with strongly correlated
electrons is evidently a difficult subject. In spite of its relevance
for the intensively investigated anomalous properties of cuprates,
there has been so far no consensus on the behavior and even
less in the appropriate formalism for an analytical evaluation of the
$R_H(\omega)$. 

It is clearly an advantage to deal with the system at $T=0$ since here
the transport (reactive) Hall constant $R_H^0$ is well defined but
does not involve any scattering or dissipation. One could hope that
such $R_H^0$ remains a reasonable approximation for $R_H(T)$ at finite
but small $T>0$. This is for example the case for normal metals and
semiconductors where within the approximation of uniform (in ${\bf
k}$) but $T$ dependent relaxation rate $\tau(T)$ the latter cancels
out and finally $R_H(T)\sim R_H^0$.

The central quantity in our approach for $R_H(\omega)$ is the
off-diagonal stiffness $\Lambda_{yx}$ which plays analogous role as
the charge stiffness $D_{\alpha\alpha}$ in the diagonal optical
conductivity $\sigma_{\alpha\alpha}(\omega)$.  We show in Sec. II that
$\Lambda_{yx}$ can be expressed in terms of matrix elements involving
solely lowest excited state (at $B=0$) which is a conceptual and
technical simplification, which is also well adapted for application
to a broader class of Fermi liquid systems. For a Fermi system with a
well defined Fermi surface and gapless electron-hole excitations we
also find a formal correspondence (apart from some ambiguities with
the second term in Eq.(\ref{eq45})) of the expression for
$\Lambda_{yx}$ with the one in the relaxation-time approximation. The
main difference in correlated system is that effective quantities
${\bf v}({\bf k})$ and ${\bf j}({\bf k})$ are not directly related.

From the general formalism in Sec. II it is clear that there is
intimate relation between $\Lambda_{yx}$ and the matrix elements of the
stiffness operator, Eqs.(\ref{eq22}),(\ref{eq22a}). More directly we can
express $R_H^0$ with the $D(n)$ itself in the case of quasi-1D
correlated system, where we recover (under certain restrictions) the
expression (\ref{eq1}) derived quite generally in the slow limit
(first $\omega=0$, then $q\to 0$).  The Hall response of such quasi-1D
systems is of direct experimental interest, in particular recently 
in connection with the apparent controversies in 1D conductors
\cite{miha} as well as with the striking vanishing of the Hall constant 
in the stripe phase of cuprates \cite{uchi,prel2}. Since strongly
correlated quasi-1D systems are not expected to be singular
one can expect that the relation (\ref{eq1}) remains qualitatively
valid even for a broader class of strongly correlated electrons.

\begin{acknowledgments}

P.P. acknowledges the support of the
Institute for Theoretical Physics, ETH Z\"urich, IRRMA and EPFL
Lausanne, where parts of this work has been done, as well as the
support of the Ministry of Education, Science and Sport of Slovenia.
X.Z. acknowledges the support by the Swiss National Science
Foundation, the University of Fribourg and the University of
Neuch\^atel.

\end{acknowledgments}

\end{document}